\pdfoutput=1
\documentclass[superscriptaddress, 12pt, aps, prl,  preprint, longbibliography]{revtex4-1}
\usepackage{graphicx}
\usepackage{epstopdf}
\usepackage{amsmath, amssymb}
\usepackage{color, soul}
\DeclareUnicodeCharacter{0308}{\"{u}}
\DeclareUnicodeCharacter{030C}{\v{ }}
\soulregister\cite7

\begin{document}

\title{Synthesis and stability of biomolecules in C-H-O-N fluids under Earth's upper mantle conditions}

\author{
Tao Li$^{1}$, Nore Stolte$^{1,a}$, Renbiao Tao$^{2}$, Dimitri A. Sverjensky$^{3}$, Isabelle Daniel$^{4}$, Ding Pan$^{1,5,*}$ \newline
        \small $^{1}$\textit{Department of Physics, Hong Kong University of Science and Technology, Hong Kong 999077, China} \newline
        \small $^{2}$\textit{Center for High Pressure Science and Technology Advanced Research (HPSTAR), Beijing 100193, China} \newline
        \small $^{3}$\textit{Department of Earth and Planetary Sciences, Johns Hopkins University, 3400 North Charles Street, Baltimore, Maryland 21218, United States} \newline
        \small $^{4}$\textit{Universite Claude Bernard Lyon1, LGL-TPE, UMR 5276, CNRS, Ens de Lyon, Universite Jean Monnet Saint-Etienne, Villeurbanne, 69622, France} \newline
        \small $^{5}$\textit{Department of Chemistry, Hong Kong University of Science and Technology, Hong Kong 999077, China} \newline    
        \small $^{a}$\textbf{Present address: \# N.S.: Lehrstuhl f{\"u}r Theoretische Chemie, Ruhr-Universit{\"a}t Bochum, 44780 Bochum, Germany} \newline
        \small $^{*}$\textbf{Corresponding author: Ding Pan, Email: \tt{dingpan@ust.hk}}       
}

\date{\today}
\begin{abstract}
How life started on Earth is an unsolved mystery. There are various hypotheses for the location ranging from outer space to the seafloor, subseafloor or potentially deeper. Here, we applied extensive ab initio molecular dynamics (AIMD) simulations  to study chemical reactions between NH$_3$, H$_2$O, H$_2$, and CO  at pressures (P) and temperatures (T) approximating the conditions of Earth’s upper mantle  (\emph{i.e.} 10 - 13 GPa, 1000 - 1400 K).
Contrary to the previous assumptions that large organic molecules might readily disintegrate in aqueous solutions at extreme P-T conditions, we found that many organic compounds 
formed  without any catalysts and persisted in C-H-O-N fluids under these extreme conditions, including  glycine, ribose, urea, and uracil-like molecules. Particularly, our free energy calculations showed that the C-N bond is thermodynamically stable at 10 GPa and 1400 K. 
Moreover, while the pyranose (six-membered-ring) form of ribose is more stable than the furanose (five-membered-ring) form at ambient conditions, 
we found that the formation of the five-membered-ring form of ribose is thermodynamically more favored at extreme conditions, which is consistent with the exclusive incorporation of $\beta$-D-ribofuranose in RNA.
We have uncovered a previously unexplored pathway through which the crucial biomolecules could be abiotically synthesized from geofluids in the deep interior of Earth and other planets and these formed biomolecules could potentially contribute to the early stage of the emergency of life.

\end{abstract}
\maketitle

\section{Introduction}
The origin of life on Earth remains a profound scientific mystery.
Numerous hypotheses have been put forward, but not a single one can explain all \cite{orgel1994origin, mann2013origins}. 
As early as $\sim$150 years ago, Darwin proposed a ``warm little pond" idea  \cite{darwin1871jd}, which was later developed by Oparin and Haldane into the influential ``primordial soup" theory in the 1920s \cite{oparin1957origin}, that is, the reactions of small molecules, such as CH$_4$, NH$_3$, H$_2$O, and CO$_2$, to form the first organic compounds. This leads to the abiotic origin of life on early Earth. 
In 1953, the famous Miller-Urey experiment was conducted to test this hypothesis at the presence of electric discharge \cite{miller1953production}.
Inspired by the ``primordial soup" theory, various prebiotic environments have been proposed for the origin of life, ranging from the outer space \cite{ziurys2006chemistry,  
sandford2020prebiotic,
krasnokutski2022pathway}
to deep-sea hydrothermal vents \cite{daniel2006origins, martin2008hydrothermal}.

In those hypotheses, prebiotic chemistry typically happens at non-ambient conditions.
The pressure (P) of hydrothermal vents, resulting from the interaction between seafloor and seawater producing hot fluids, is tens of megapascals depending on the depth of water, and the temperature (T) can reach 700 K. Those deep sea vents 
were considered to have suitable conditions for the origin of life \cite{martin2008hydrothermal}, 
but previous studies also suggested that the high temperatures in the vents could readily degrade crucial biomolecules in aqueous solutions \cite{miller1988submarine}. 
Shock recovery experiments successfully synthesized various biomolecules, where the pressure can reach tens of gigapascals and the temperature can rise to thousands of kelvins before releasing back to ambient conditions \cite{Sekine2024}. However, due to the dynamic nature of shock waves, samples do not reach thermodynamic equilibrium states, so the shock experiments are usually conducted to mimic the conditions of meteorite impacts on Earth \cite{furukawa2009biomolecule}.
It was commonly assumed that large biomolecules could not persist at both high hydrostatic pressures and high temperatures as found in Earth's upper mantle, where pressures can reach up to $\sim$13 GPa,
accompanied by high temperatures of $\sim$1700 K.
In recent years, Otake et al. studied the stability of amino acids and their oligomerization at 
1.0--5.5 GPa
and 453--673 K, and found that high-pressure conditions inhibited the decomposition of amino acids and small amino acids could be oligomerized up to
pentamers. However, their experiments were conducted at the water-poor condition \cite{otake2011stability}.
In aqueous solutions, there seems a strong thermodynamic drive towards the formation of amino acids dimers above 1.5 GPa and temperature above 423 K \cite{robinson2021quantifying}, in agreement with some experimental observations \cite{pedreira2019spontaneous}.  
 Previously, the fluids in Earth's mantle were commonly modeled as simple mixtures of small unreactive volatile molecules, including H$_2$O, CO$_2$, CO, CH$_4$, and H$_2$ \cite{zhang2009model}. However, recent experimental and theoretical studies have revealed the significant roles played by chemical speciation, aqueous ions, and complexes in the supercritical geofluids within Earth's lithosphere \cite{manning2013chemistry,pan2013dielectric, sverjensky2020changing, pan2016fate, abramson2017water, stolte2019large, stolte2021water}. These findings have challenged the previous assumptions and shed light on the potential existence of larger organic molecules in Earth's or other planets' subsurface or even deeper.

It is very challenging to study the building blocks of biomolecules for life in laboratory settings due to the extreme environments and subsequent analyses.
Atomistic simulations based on first principles have been applied to study prebiotic chemistry, offering significant molecular-level insights \cite{perez2020prebiotic}.
Saitta \emph{et al.} applied ab initio molecular dynamics (AIMD) and metadynamics simulations to study the Miller-Urey experiment, and found unexpected reaction intermediates including formic acid and formamide \cite{saitta2014miller}. 
Goldman \emph{et al.} applied AIMD simulations to investigate the formation of prebiotic compounds from impact-induced shock compression of cometary ices followed by expansion to ambient conditions, and found that the impact of cometary ices could produce amino acids independently of atmospheric conditions and Earth minerals \cite{goldman2010synthesis, goldman2013prebiotic, koziol2015prebiotic}.
Note that the shock compression simulations did not reach thermal equilibrium, and 
it is unknown whether the biomolecules can remain stable without being quenched to ambient conditions \cite{Ross2013comment, Goldman2013reply}.
AIMD simulations coupled with enhanced sampling techniques have also been applied to study the role of mineral surfaces in prebiotic chemistry \cite{nair2006glycine,pollet2006ab, schreiner2011peptide}.
Despite the numerous previous studies, the exploration of prebiotic chemistry at both high hydrostatic pressures and high temperatures  largely remains an uncharted territory.

Here, we applied extensive AIMD simulations to study the reactions of inorganic C-H-O-N geofluids at 10--13 GPa and 1000--1400 K, the P-T conditions as found in the transition zone of Earth's upper mantle. 
We discovered that a large range of organic molecules directly formed without any catalysts at extreme P-T conditions, including formamide, glycine, ribose, urea, and uracil-like molecules. The formation pathways are also detailed discussed. 
Our free energy calculations using AIMD simulations suggest that the C-N bond is thermodynamically stable at 10 GPa and 1400 K. Our study suggests a new potential avenue for the abiotic formation of biomolecules in the deep interior of Earth and other planets, further expanding the applicable places of ``primordial soup" theory.

\section{Results and Discussion}
\begin{figure}
\centering
\includegraphics[width=1.0\textwidth]{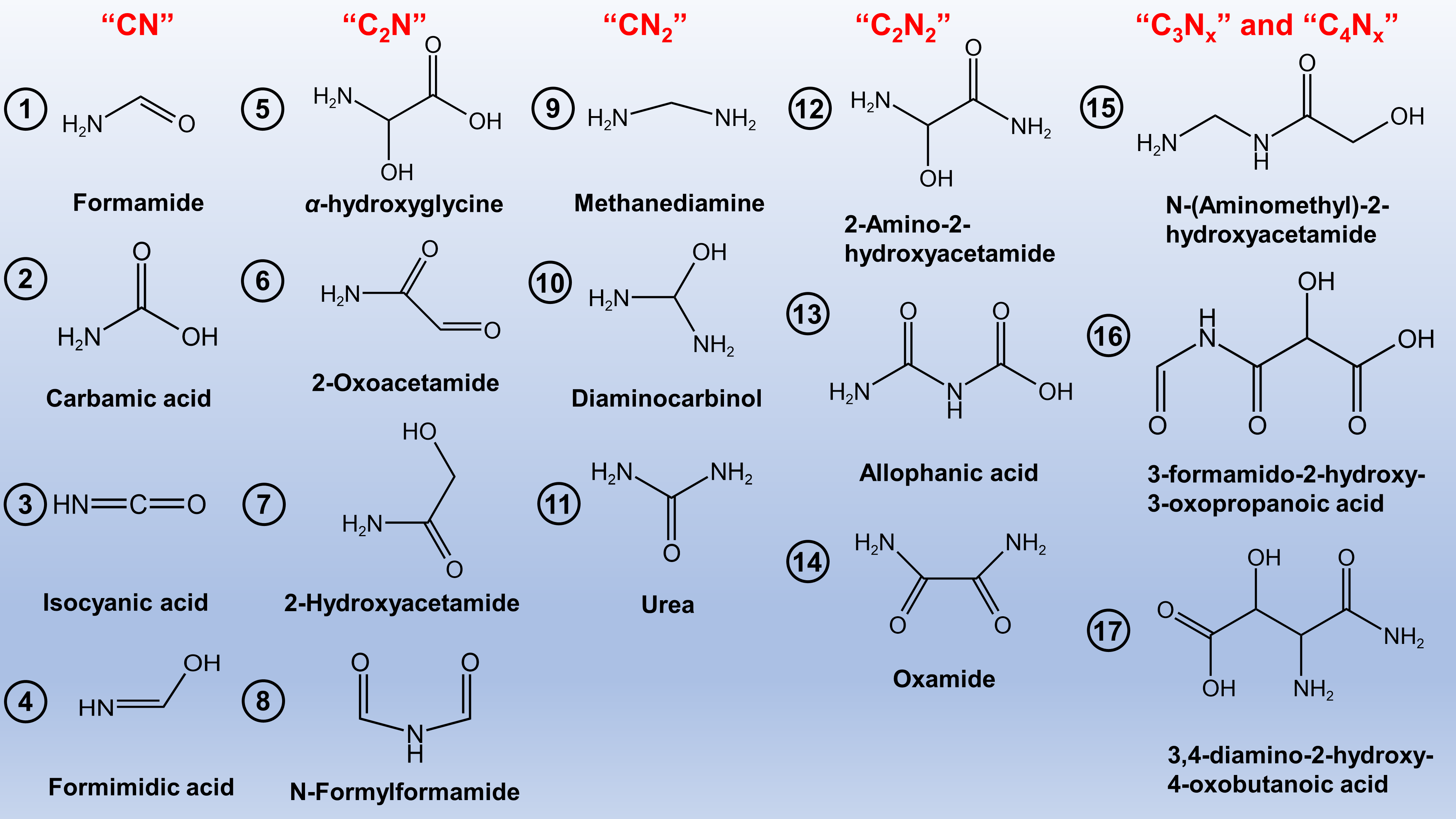}
\caption{Chemical structures of the representative products for the “CN”, “C$_2$N”, “CN$_2$”, “C$_2$N$_2$”, “C$_3$N$_x$” and “C$_4$N$_x$” species.}
\label{Fig:chem_struc}
\end{figure}

The simulation box initially contained 15 molecules of each of the following species:
H$_2$O, H$_2$, CO, and NH$_3$ (Fig. S1). 
Those molecules are typical volatile compounds found in C-H-O-N fluids in reducing environments of Earth's deep interior \cite{zhang2009model, manning2013chemistry}, and are also abundant in interstellar space \cite{guelin2022organic}.
We conducted AIMD simulations at three P-T conditions, 10 GPa and 1000 K, 10 GPa and 1400 K, and 13 GPa and 1400 K, to evaluate how chemical products vary with P and T. Those P-T conditions are typically found at the bottom of Earth's upper mantle.
After more than 400 ps simulations at each P-T condition, we found about 100 different organic species.
Here, we mainly focus on the formation of biotically relevant molecules, so we classify the organic molecules into six categories: 
``CN", ``C$_2$N", ``CN$_2$", ``C$_2$N$_2$", ``C$_3$N$_x$", and ``C$_4$N$_x$", based on the numbers of carbon and nitrogen atoms, as shown in Table \ref{Tab_CN}.
Fig. \ref{Fig:chem_struc} shows the chemical structures of the representative species. 
Species with more than five carbon atoms were observed, however they are  excluded at this stage due to their high instability and extremely short lifetime, typically lasting for only a few femtoseconds.  The percents of these CN-containing species as a function of time at three different P-T conditions are shown in Fig. S3, which suggest that the concentrations of these species gradually stabilize over time. 

\begin{figure}
\centering
\includegraphics[width=1.0\textwidth]{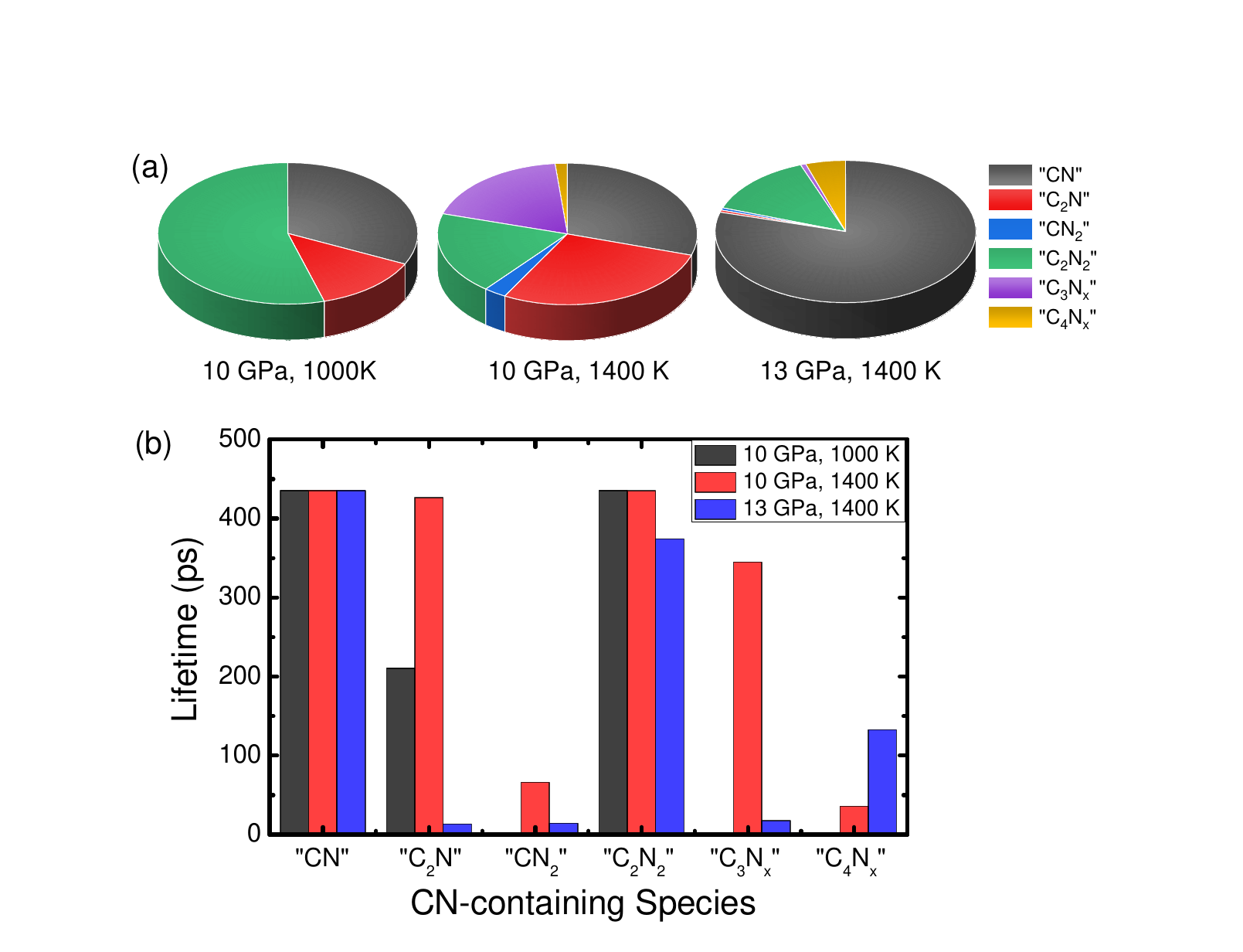}
\caption{ CN-containing species at three P-T conditions. 
\textbf{(a)} Fractions and \textbf{(b)} average lifetime of the six CN-containing species obtained from AIMD simulations.}
\label{Fig:distri_CN}
\end{figure}

At 10 GPa and 1000 K, most of the organic products belong to three main categories, as shown in Fig. \ref{Fig:distri_CN}(a).
The ``C$_2$N$_2$" species accounts for more than half of the CN-containing species, followed by the ``CN" species, and finally ``C$_2$N" species. 
Formamide is the primary product in the ``CN" species \cite{saladino2012formamide}. It contains the peptide bond, which is a significant precursor for the production of more complex biomolecules, such as amino acids \cite{pietrucci2015formamide}, nucleic acids, proteins, and even sugars \cite{saladino2012genetics,halfen2011formation}. The ``C$_2$N" species has the same CN-backbone \mbox{(``-N-C-C-")} as glycine, suggesting that the ``C$_2$N" species may be a potential reactant for the formation of amino acids. At this P-T condition, some of the ``C$_2$N" species are not stable and can be transformed into the ``C$_2$N$_2$" species by reacting with ammonia or ammonium ions, so the CN-backbone of ``C$_2$N$_2$" species is ``-N-C-C-N-", and one interesting molecule is identified as alpha-hydroxy-glycineamide ($\alpha$-HGA).
 
With increasing the temperature to 1400 K along the isobar at 10 GPa, all the six categories of organic products at this P-T condition were observed. Fractions of the ``CN$_2$" and ``C$_4$N$_x$" species are very small, while the other four species have similar fractions. Compared with 10 GPa and 1000 K, the portion of ``C$_2$N$_2$" species becomes smaller, while there are more ``C$_2$N" species. The ``C$_2$N" species formed here also shares the same CN-backbone (``-N-C-C-") with glycine. Notably, in the ``C$_2$N" species, we found a molecule of $\alpha$-hydroxyglycine, which can easily evolve into glycine \cite{saitta2014miller}. The ``C$_2$N$_2$" species has the CN-backbone of ``-N-C-C-N-", including $\alpha$-HGA and oxamide. They are important organic molecules in the formation of the building blocks of peptides \cite{ligterink2019formation}. Among the ``CN" species, formamide remains the most abundant one. Additionally, the ``CN$_2$" species with the ``-N-C-N-" backbone was observed, which did not typically form at 10 GPa and 1000 K.  One representative product of this species is methanediamine (CH$_6$N$_2$) \cite{carballeira2001role}, which was reported as a vital intermediate or precursor in the abiotic formation of nucleobases \cite{marks2022preparation}.

With increasing the pressure to 13 GPa at 1400 K,
the ``CN" species becomes dominating, and the ``C$_2$N$_2$" species comes next. For the ``CN" species, besides formamide, some other crucial precursors for producing more complex biomolecules are identified, such as carbamic acid \cite{tossell2010happens}, isocyanic acid \cite{belson1982preparation}, and formimidic acid \cite{pranata1995computational}. The ``C$_2$N$_2$" species exhibits a distinct CN-backbone compared to the two solutions at 10 GPa, forming as ``-C-N-C-N-", which is present in allophanic acid. This acid plays a vital role in the synthesis of biomolecules, \emph{e.g.}, uracil, and would be used as reactants in subsequent simulations of forming larger biomolecules. Surprisingly, the fraction of ``C$_2$N"  species sharply decreases, adopting a new CN-backbone structure of ``-C-N-C-", as observed in N-formylformamide \cite{sanz2009effects}. This structural change makes it challenging for ``C$_2$N"  species formed here to evolve into glycine-like molecules or larger amino acids. In the ``CN$_2$" species, we discovered the presence of urea molecule. Urea is closely related with biological processes \cite{daniel2006origins} and considered to be an important condensation agent in prebiotic chemistry \cite{burcar2019stark, gull2020silicate}, so it would also serve as a valuable reactant for investigating the reactions to larger biomolecules.  

For all the three investigated P-T conditions, a common feature is that the ``CN" and ``C$_2$N$_2$" species are always the major components among the six CN-containing species.  
The lifetime distribution in Fig. \ref{Fig:distri_CN} (b) shows that 
 the ``CN" and ``C$_2$N$_2$" species could exist for a longer time at all three P-T conditions, suggesting that they are relatively more stable than other species.

\begin{figure}
\centering
\includegraphics[width=1.0\textwidth]{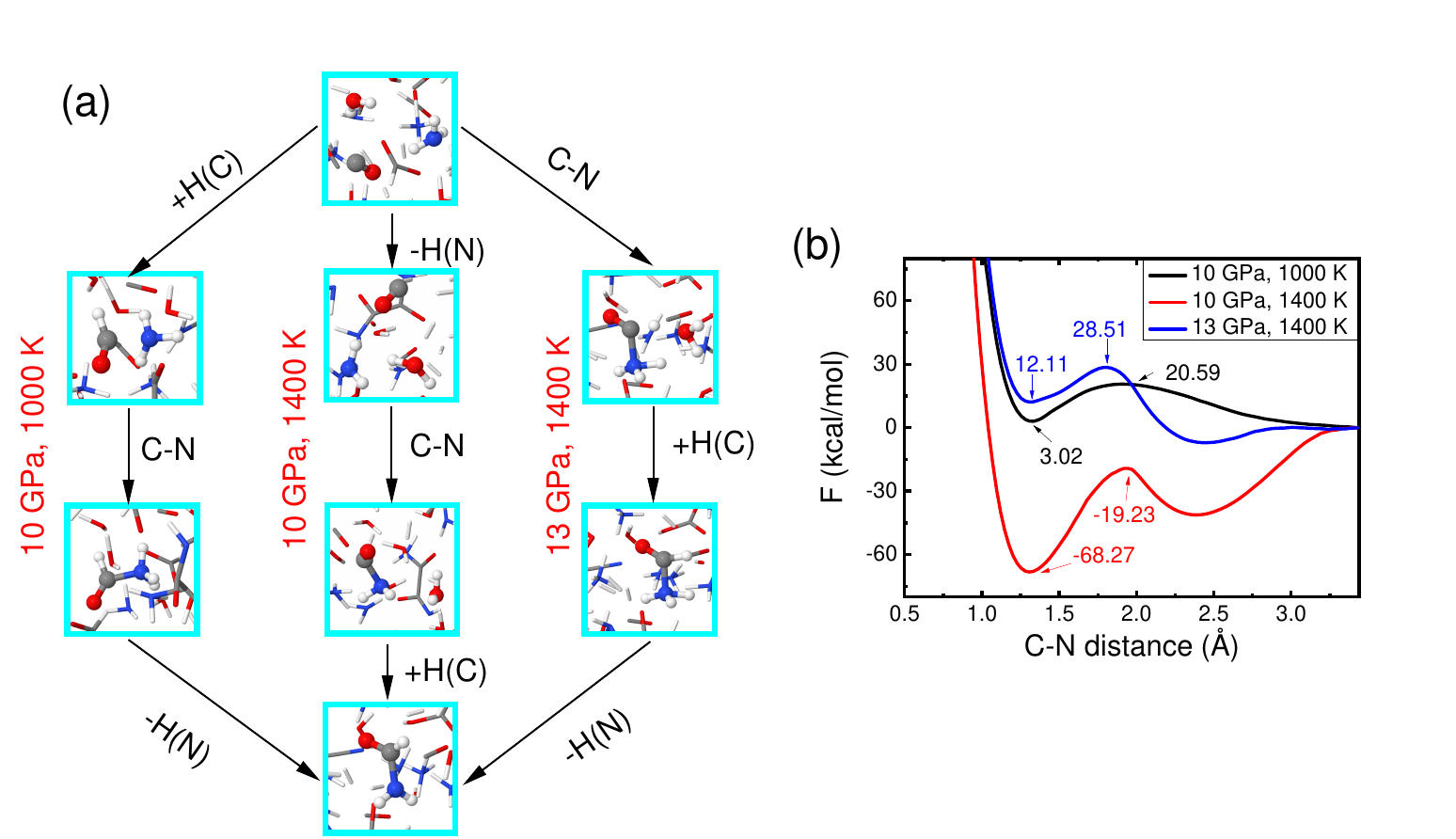}
\caption{Synthesis of formamide from H$_2$O, H$_2$, CO, and NH$_3$ at three P-T conditions.
There were initially 15 molecules for each of these four small molecules in the simulation box.
\textbf{(a)} The key reaction steps. The labels over arrows, +H(C), C-N, and -H(N), refer to the formation of a C-H bond, a C-N bond, and the breaking of a N-H bond, respectively. \textbf{(b)} The free energies obtained at three P-T condtions as functions of the C-N distance.}
\label{Fig:free_energy_CN}
\end{figure}

Among these generated chemical products, formamide is a crucial precursor molecule that plays a vital role in the synthesis of biomolecules in prebiotic chemistry \cite{saladino2012formamide,pietrucci2015formamide,saladino2015meteorite,rios2015impact,ferus2015high, enchev2021chemical}. Thus, we first investigated the formation of formamide molecule.
Fig. \ref{Fig:free_energy_CN} (a) 
shows the different formation pathways at three P-T conditions, and detailed formation processes are displayed in Fig. S4. 
At 10 GPa and 1000 K, the formation reaction begins with the dissociation of water molecules, generating OH$^-$ and H$^+$ ions in the aqueous solution. 
The encounter between an H$^+$ ion and a CO molecule leads to the formation of a C-H bond and an aldehyde group (-CH=O).
Subsequently, the aldehyde group reacts with a NH$_3$ molecule to generate a C-N bond, and then, one N-H bond breaks, ultimately producing formamide.
At 10 GPa and 1400 K, the formation process differs from that at 10 GPa and 1000 K. One N-H bond in NH$_3$ first breaks, leading to  a dissociative amino group (-NH$_2$). The amino group then reacts with a CO molecule, forming the C-N bond. Later, a water molecule dissociates, providing a proton that ultimately contributes to the formation of the formamide molecule. 
At 13 GPa and 1400 K, the C-N bond is initially formed through the reaction between NH$_3$ and CO molecules. After a very short time, a dissociated proton from a water molecule is also bonded to the carbon atom, forming a C-H bond. Finally, by releasing a proton from the NH$_3$ group, the formamide molecule is generated. 
Despite the different formation pathways at three P-T conditions, we observed two common features: first, the reactants always have CO, H$_2$O, and NH$_3$, and second, the reactions must involve the formation of C-N and C-H bonds and the breakage of N-H bonds.

For the generation of biomolecules in living matter, the formation of C-N bonds is the key step. 
We further applied the advanced sampling method (see details in the Computational Methodology section) to calculate the free energy profile of the C-N bond formation at three P-T conditions in Fig. \ref{Fig:free_energy_CN}(b).
The collective variable (CV) is the distance between the carbon and nitrogen atoms.
Our calculations show that the C-N bonds at 10 GPa, 1000 K and 13 GPa, 1400 K, whose lengths are 1.33 \AA\ and  1.31  \AA\, respectively, are at the metastable states. 
The free energy decreases upon dissociation of the C-N bond,
indicating that breaking C-N bonds at these two P-T conditions are more thermodynamically favored. 
The free energy profiles suggest that the energy barriers of the C-N bond dissociation are
17.57 kcal/mol and 16.40 kcal/mol, 
at 10 GPa, 1000 K and 13 GPa, 1400 K, respectively.
As a comparison, at 10 GPa and 1400 K, 
the free energy increases upon dissociation of the C-N bond (1.31 \AA), and the energy barrier is 49.04 kcal/mol, indicating that the C-N bond is thermodynamically and kinetically stable at this P-T condition.  
The free energy result agrees with the data in Fig. \ref{Fig:distri_CN}, where  there are more CN-containing species at 10 GPa and 1400 K than at the other two P-T conditions. Additionally, compared with the free energy profile at 10 GPa and 1000 K, there 
are two more energy minima at the C-N distance of 2.37 \AA\  and 2.45 \AA\ at 10 GPa, 1400 K and 13 GPa, 1400 K, respectively. The free energy increases upon the dissociation of the C-N pair, indicating that these two minima are thermodynamically stable. The C-N distances are larger than that of a typical C-N covalent bond,  which implies the presence of an additional atom connecting the two.
This is consistent with a wider variety of organic species at these two P-T conditions. 
Overall, our free energy calculations suggest that
the C-N bond formation is favored at 10 GPa and 1400 K.

\begin{figure}
\centering
\includegraphics[width=1.0\textwidth]{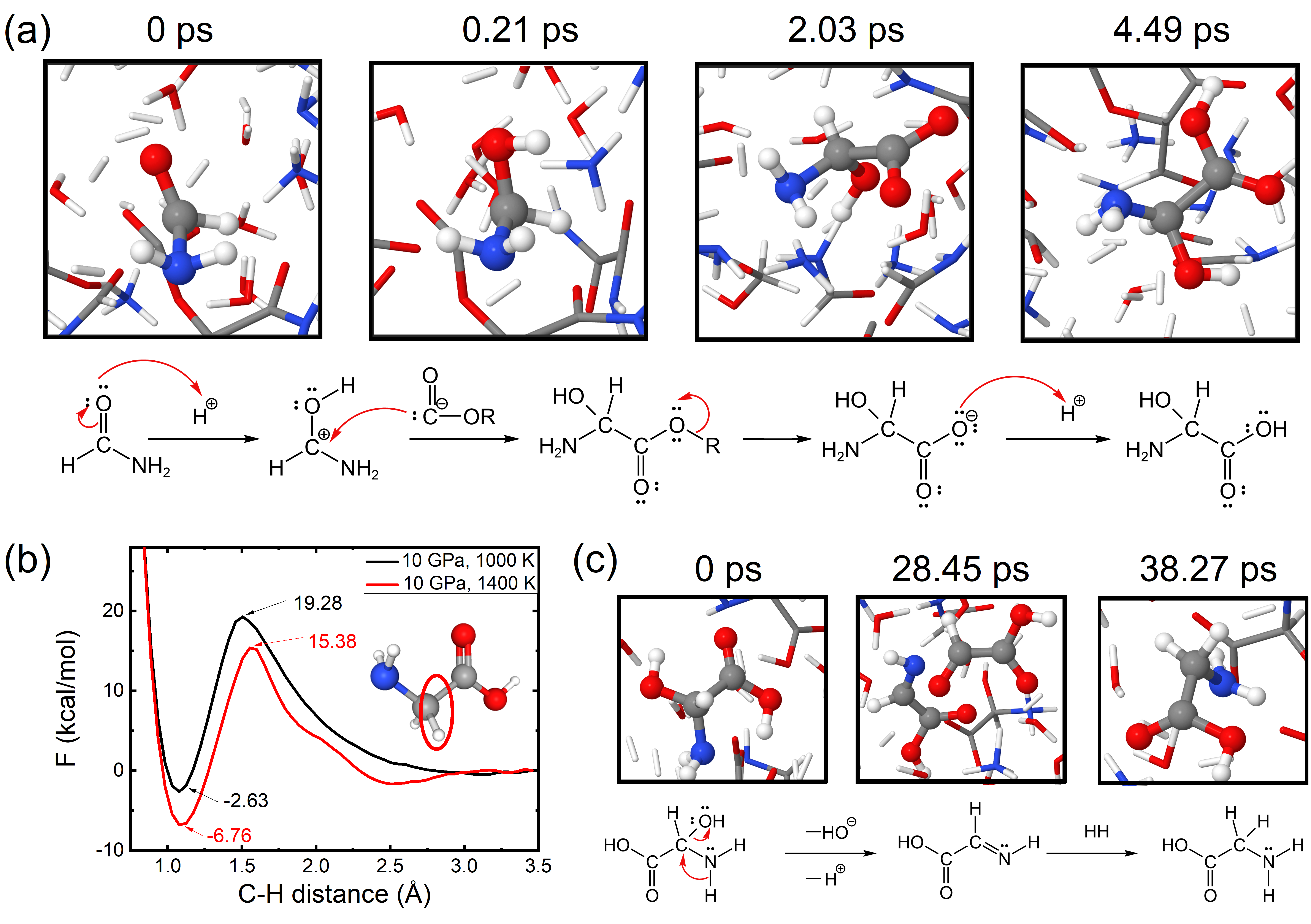}
\caption{Synthesis of $\alpha$-hydroxyglycine and glycine at 10 GPa and 1400 K.
\textbf{(a)} Key reaction steps and reaction mechanism in the formation of $\alpha$-hydroxyglycine from formamide. 
\textbf{(b)} Free energies at two P-T conditions as functions of the C-H distance. \textbf{(c)} Key reaction steps and reaction mechanism in the formation of glycine from 5 $\alpha$-hydroxyglycine, 10 H$_2$, and 10 H$_2$O. 
The curved arrows show the movement of electrons.} 
\label{Fig:alpha-hydroxyglycine}
\end{figure}

After studying the basic chemical products at three P-T conditions, we now turn to investigate the formation of large biomolecules as potential building blocks of life. Previous simulations have extensively studied glycine and the smallest amino acid under various conditions \cite{saitta2014miller,goldman2010synthesis,wang2014discovering}. In our simulations at 10 GPa and 1400 K, we discovered  $\alpha$-hydroxyglycine, very similar to glycine. Fig. \ref{Fig:alpha-hydroxyglycine} (a) illustrates the process of forming $\alpha$-hydroxyglycine from formamide.  A proton attacks the formamide molecule, leading to the creation of a protonated formamide molecule. Then, the protonated formamide reacts with a large CO$_2$-containing molecule, resulting in the ``-N-C-C-" backbone through forming a C-C bond.
Finally, another proton is captured by a “-COO” group, ultimately yielding the $\alpha$-hydroxyglycine molecule. In comparison to glycine, $\alpha$-hydroxyglycine differs by having one hydrogen atom and an “-OH” group instead of two H atoms connecting to $\alpha$-C. Glycine can be obtained if the “-OH” group is replaced by a  hydrogen atom. During this process, the formation of the C-H bond is a crucial step. 
To explore appropriate P-T conditions for the possible synthesis of glycine, 
we calculated the free energy as a function of the C-H distance in Fig. \ref{Fig:alpha-hydroxyglycine} (b). Considering that the CN-backbone of the ``C$_2$N" species at 13 GPa - 1400 K is ``-C-N-C-",  different from the CN-backbone in the glycine molecule, we did not consider this P-T condition.  We can see that the free energy minima is at 1.07 \AA\ for both 10 GPa, 1000 K and 10 GPa, 1400 K, and the energy barriers to break the C-H bond are 21.91 kcal/mol and 22.14 kcal/mol, respectively, which are very similar. 
After the C-H bond dissociation, the free energy increases more at 10 GPa and 1400 K (6.76 kcal/mol) than at 10 GPa and 1000 K (2.63 kcal/mol), indicating that the C-H bond is more favored at the former P-T condition.

To study the reaction from $\alpha$-hydroxyglycine to glycine at extreme P-T conditions, we set up a new simulation box containing 5 $\alpha$-hydroxyglycine, 10 H$_2$, and 10 H$_2$O molecules, as shown in Fig. \ref{Fig:alpha-hydroxyglycine}(c) and performed AIMD simulations at 10 GPa and 1400 K. Initially, two $\alpha$-hydroxyglycine molecules combine to form one molecule, but later dissociate  
to  dehydroglycine and glyoxylic acid after $\sim$28 ps. 
As the simulation goes on, the dehydroglycine will be reduced by H$_2$ and finally it becomes glycine at $\sim$ 38 ps. 

Both proteins and DNA are widely recognized as the important biomolecules for building blocks of life. However, the question of which one came first in the emergence of life remains a topic of debate. This is due to the fact that the replication of DNA relies on the catalytic activity of enzymes, which are a type of proteins. Conversely, the production of proteins also needs the guidance of DNA, resulting in a chicken-and-egg dilemma. To solve this problem, a hypothesis known as “the RNA world” was proposed, which believed that RNA may occur earlier than DNA and protein \cite{gilbert1986origin,nisbet1986origin}. During this stage, the RNA not only have the ability to reproduce themselves and carry genetic information, but also could catalyze certain biological reactions in life activities like enzymes \cite{mcclain1987model,forster1990external}. Thus, the synthesis of RNA plays an critical role in the early stage of the origin of life. RNA is composed of ribose sugar, four kinds of nucleobases, and a phosphate group. Considering that our simulations only involve the
carbon, hydrogen, oxygen, and nitrogen elements, we then explored whether the ribose and nucleobases could be synthesized under extreme conditions.

\begin{figure}
\centering
\includegraphics[width=1.0\textwidth]{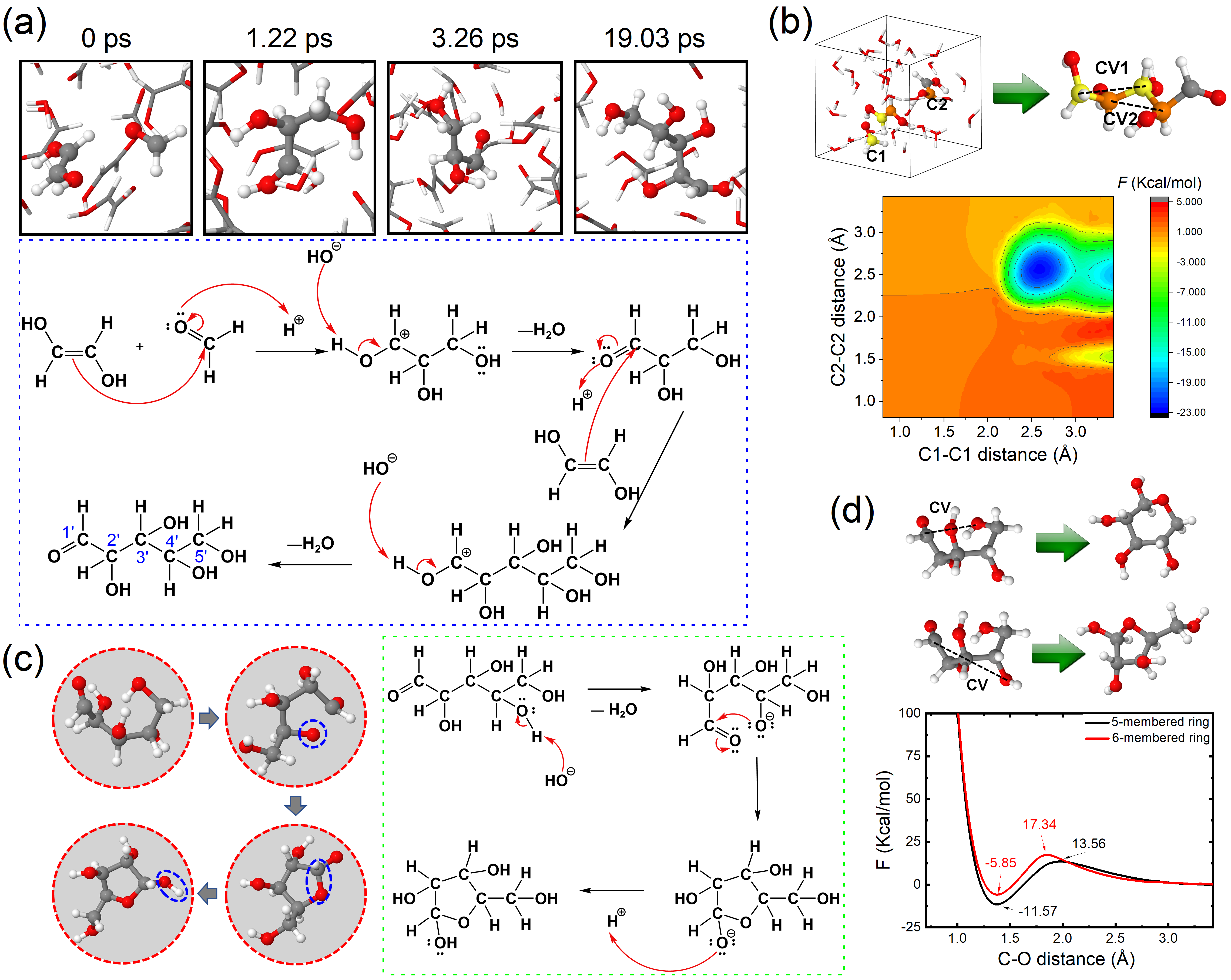}
\caption{Synthesis of ribose at 10 GPa and 1400 K.
\textbf{(a)} Key reaction steps and reaction mechanism in the formation of the open-chain ribose from 10 (Z)-ethene-1,2-diol, 5 formaldehyde, and 5 H$_2$O. \textbf{(b)} The free energy landscape for the formation of ribose. Two C-C distances are the collective variables, CV1, and CV2. To distinguish the selected carbon atoms, C1 and C2 are marked by yellow and orange colors, respectively. \textbf{(c)} The transformation of ribose from open chain to five-membered ring and related reaction mechanism. \textbf{(d)} Free energies of the formation of
furanose (five-membered ring) and pyranose (six-membered ring) forms of ribose. The C-O distance is the collective variable.The curved arrows show the movement of electrons.}
\label{Fig:ribose}
\end{figure}

We first considered ribose (C$_5$H$_{10}$O$_5$). The formose reaction is a well-known prebiotic pathway for synthesizing ribose from formaldehyde (CH$_2$O) \cite{breslow1959mechanism}. Our simulations have generated a few relevant hydrocarbon products (shown in Figs. S5 and S6), including formaldehyde molecule (number 1 in Fig. S5),  which was later chosen as the primary reactant. Additionally, to facilitate the synthesis process, the reactants also included a “C$_2$” molecule, (Z)-ethene-1,2-diol (C$_2$H$_4$O$_2$, number 7 in Fig. S5). Initially, we constructed a simulation box consisting of 10 (Z)-ethene-1,2-diol molecules and 5 formaldehyde molecules (Fig. S7). After $\sim$76 ps, a “C$_5$” molecule was formed at $\sim$10 GPa and 1400 K. However, it did not transform into the structure of ribose after an extended time. 
We further tried to add other reactants to the simulation box, and found that after adding 5 water molecules, the ribose molecule appeared, indicating the importance role of water in the formation of ribose. The reaction process is shown in Fig. \ref{Fig:ribose} (a), where the formaldehyde and “C$_2$” molecules react to form a “C$_3$” molecule, which then combines with another “C$_2$” molecule to form a “C$_5$” molecule. This “C$_5$” molecule then evolves into the open-chain form of ribose at $\sim$19 ps.

We calculated the free energy landscape in Fig. \ref{Fig:ribose} (b), where two C-C distances were selected as collective variables labelled as CV1 and CV2.  
For CV1, two carbon atoms come from formaldehyde and a “C$_2$” molecule, respectively, which are labeled as C1. While for CV2, two carbon atoms are from two different “C$_2$” molecules, marked as C2. 
The free energy landscape shows a clear energy minimum corresponding to the formation of ribose, but the C-C distance at this minimum is larger than a typical C-C bond distance.
We examined the biased molecular dynamics trajectories and found that the carbon backbone of the generated ribose molecule has the  C1-C2-C1-C2 structure. 
By measuring the distances of C1-C1 and C2-C2 , we confirmed that they are consistent to the distances observed in the free energy landscape. 
Furthermore, another initial configuration for the free energy calculation is also presented in Fig. S8, the obtained free energy landscape still shows that the ribose molecule is thermodynamically stable.

Previous studies showed that the cyclic forms of ribose are abundant in aqueous solutions at room temperature \cite{drew199813c}. However, the cyclic forms of ribose, either furanose form (five-membered ring) or pyranose form (six-membered ring),  were not detected in our simulations (Fig. \ref{Fig:ribose} (a)). 
The configuration change of ribose may require very long simulation time. 
To further study the cyclic formation mechanism, 
we tried a few additional simulations with new initial states, and finally chose the curved open-chain ribose molecules obtained from our previous simulations, as it appears that the C4' or C5' hydroxyl group has a greater chance of reacting with the C1' aldehyde group under pressure.
We built a new simulation box containing three of these ribose molecules together with 18 water molecules. After about 36 ps, 
we observed the direct formation of a furanose form of ribose.
Fig. \ref{Fig:ribose} (c) shows the mechanism of ring formation. 
In this process, the C4' hydroxyl group first loses a hydrogen atom, leaving a carbonyl group. The oxygen atom then attacks C1' to form the ring, and finally, the carbonyl group on C1' receives a dissolved proton in the solution to generate the five-membered ribose ring. Interestingly, no six-membered ring was found in our simulations at 10 GPa and 1400 K.

We further applied free energy calculations to
study the formation ability of five- and six-membered ribose rings at $\sim$ 10 GPa and 1400 K.
We chose two C-O distances as the CVs for the two rings, as shown in Fig. \ref{Fig:ribose} (d).  The energy minima are at the C-O distance of 1.37 \AA\ for the two ribose forms, indicating the formation of C-O bonds and the rings. Besides, the formation of the C-O bonds lowers the free energy by 11.57 kcal/mol for the five-membered ribose ring and 5.85 kcal/mol for the six-membered ribose ring, indicating that 
the ring formation is thermodynamically more stable and the furanose form of ribose 
is more favored than the pyranose form at $\sim$ 10 GPa and 1400 K.
The free energy result explained why the the pyranose form of ribose is absent in our AIMD simulations. The spacial limitation under high pressure may hinder the formation of larger cyclic forms of molecules,  so the energy barrier of forming a five-membered ring is lower than that of a six-membered ring.

Note that at ambient conditions, the naturally-occurring ribose exists as a mixture of cyclic forms in equilibrium with its open-chain form in aqueous solutions, and the pyranose form of ribose is the predominant conformation, accounting for approximately 70\% \cite{dewick2013essentials,dass2021equilibrium}.
However, despite the thermodynamic stability of the pyranose form of ribose over the furanose form at ambient conditions, the exclusive constituent of the carbohydrate backbone of RNA remains the furanose form of ribose. The reason behind it is still unknown. There are several hypotheses proposed to explain this preference, including silicate(borate)/ribose complexes \cite{kolb2004complexes, lambert2004silicate} and temperature gradients \cite{dass2021equilibrium}. 
Our finding suggests that extreme P-T conditions may also generate the necessary component for assembling RNA molecules. Consequently, it highlights the potential role of deep Earth environment in facilitating the emergence and development of RNA-based life.

\begin{figure}
\centering
\includegraphics[width=1.0\textwidth]{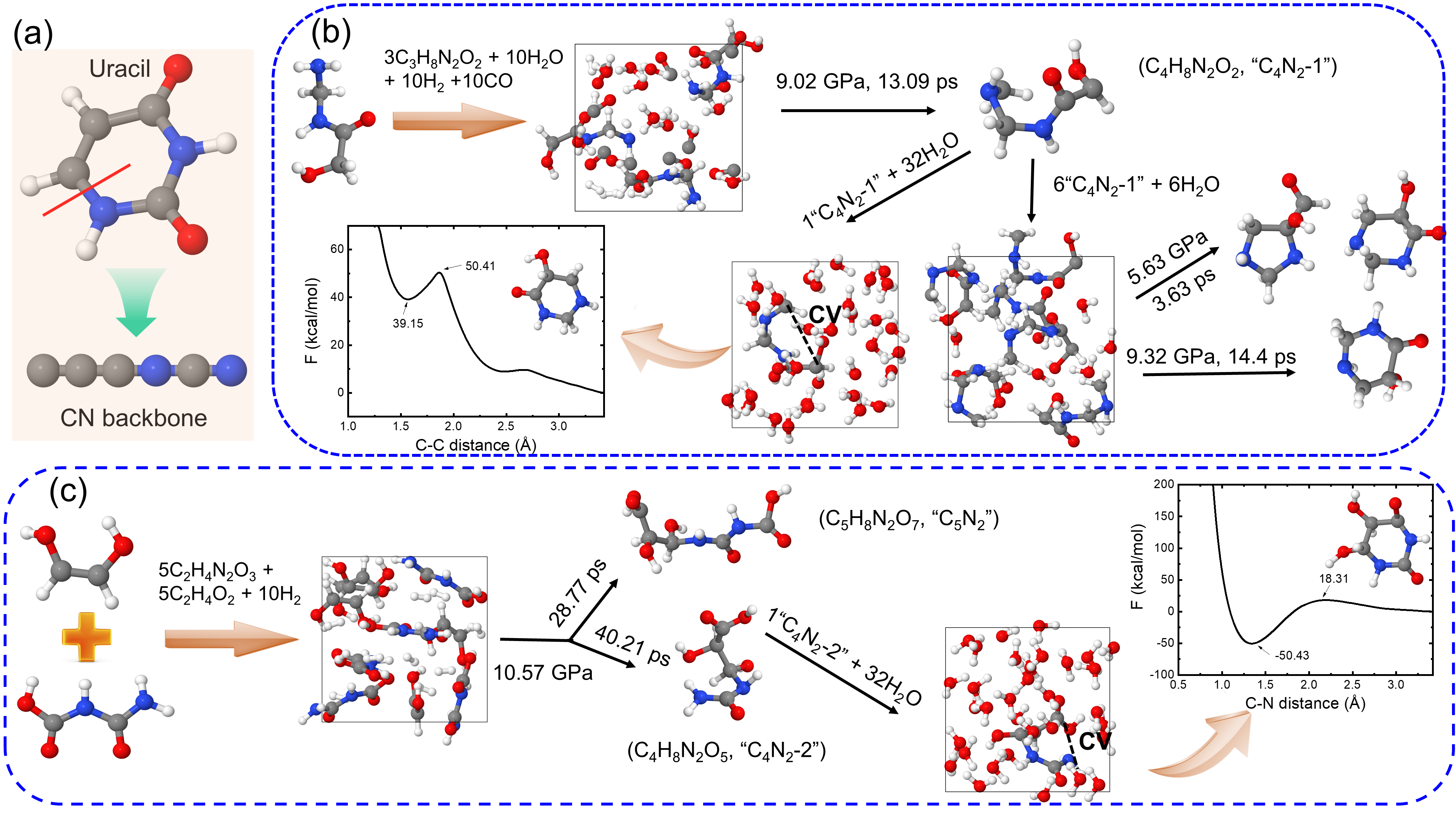}
\caption{
Synthesis of uracil-like molecules at 10 GPa and 1400 K.
\textbf{(a)} Structure of the uracil molecule and its CN-backbone when the C-N bond is broken. \textbf{(b)} The formation of the “C$_4$N$_2$-1” molecule with the same CN-backbone as the open-chain form of uracil. The reactants are 3 C$_3$H$_8$N$_2$O$_2$, 10 H$_2$O, 10 H$_2$, and 10 CO molecules in the simulation box.
We further put 6 “C$_4$N$_2$-1” and 6 H$_2$O molecules in the simulation box to study the ring formation in the uracil molecule.
The free energy calculation was conducted in the simulation box containing 
one “C$_4$N$_2$-1”, and 32 H$_2$O molecules. \textbf{(c)} 
The formation of the “C$_4$N$_2$-2” molecule with the same CN-backbone as the open-chain form of uracil.
The reactants are 5 C$_2$H$_4$N$_2$O$_3$, 5 C$_2$H$_4$O$_2$, and 10 H$_2$ molecules.
The free energy calculation was conducted in the simulation box containing one
“C$_4$N$_2$-2” and 32 H$_2$O molecules.
}
\label{Fig:urail}
\end{figure}

We then shifted our attention to the chemical evolution of nucleobases, which are another essential building blocks of genetic information. There are five naturally-occurring nucleobases: adenine (A), cytosine (C), guanine (G), thymine (T), and uracil (U). Of these, uracil is unique to RNA, so our studies primarily focused on investigating the formation of uracil molecules under extreme conditions. We first studied the formation of the open-chain “C$_4$N$_2$” molecule which has the same CN-backbone as the open-chain form of uracil, as shown in Fig. \ref{Fig:urail} (a). Previous experimental studies on the origin of life suggested that formamide is a crucial molecule for the prebiotic formation of nucleobases, in the presence of mineral or metal oxide catalysts \cite{saladino2012formamide}, heating\cite{enchev2021chemical}, and high-energy free radicals \cite{ferus2015high,ferus2014high},
so here we also first chose formamide as a reactant.
However, after about 144 ps, we did not observe the formation of the “C$_4$N$_2$” molecule at 10 GPa and 1400 K, suggesting that it may be difficult for small molecules to directly synthesize uracil in the AIMD simulations  (Fig. S9).  
Later, we chose the chemical reactants that have closer structural similarities to the “C$_4$N$_2$” molecule.
The reactant molecules, including “C$_4$N”, “C$_3$N$_2$”, “C$_2$N$_2$”, and “C$_3$”, were generated from our first-stage AIMD simulations.
A recent study suggested that uracil could be synthesized from the reactions between urea and “C$_3$” molecules \cite{choe2022mechanism}, so we specifically selected the C$_3$H$_4$O$_4$ molecule (number 10 in Fig. S5) and the urea molecule (CH$_4$N$_2$O, number 11 in Fig. \ref{Fig:chem_struc}) as reactants. The simulation results are presented in Fig. S10. However, regardless of the inclusion of water molecules or the application of different pressure conditions, we did not observe any “C$_4$N$_2$” molecules with the CN-backbone of the open-chain form of uracil.
Then we tried to find another pathways and conducted six more AIMD simulations to investigate the possibility of forming uracil through reactions involving above-mentioned molecules at various pressures and 1400 K (see Figs. S11, S12 and S14).

We found that, the C$_4$H$_8$N$_2$O$_2$ molecule, N-((($\lambda$$^3$-methyl)amino)methyl)-2-hydroxy-2$\lambda$$^3$-ethanamide (labeled as “C$_4$N$_2$-1”), which has the same CN-backbone as the open-chain form of uracil, was formed in the simulation box consisting of 3 C$_3$H$_8$N$_2$O$_2$ (number 15 in Fig. \ref{Fig:chem_struc}), 10 H$_2$O, 10 H$_2$, and 10 CO molecules, at $\sim$10 GPa and 1400 K.
The formation process is provided in Fig. S13. Like ribose, the “C$_4$N$_2$-1” molecule retained its open-chain form after an extended simulation of $\sim$108 ps.

To study the ring formation of uracil, 
we also selected the “C$_4$N$_2$-1” molecule with large curvature and built a simulation box containing 6 “C$_4$N$_2$-1” and 6 water molecules. Fig. \ref{Fig:urail} (b) shows that all the “C$_4$N$_2$-1” molecules reacted. At 5.6 GPa and 1400 K, open-chain “C$_4$N$_2$-1” molecules became five-membered and six-membered rings within a shorter time of about 3.6 ps. 
Particularly, the elements in the six-membered-ring molecule are the same as those in uracil as shown in Fig. \ref{Fig:urail}(a). At 9.3 GPa and 1400 K, the same six-membered-ring molecule was also formed. These simulations suggest that it is possible for the open-chain form of uracil to transform into the cyclic form. We further performed free energy calculations to analyze this ring formation reaction. We first chose the distance between the head and tail carbon atoms in the open-chain form of uracil as the CV. The enhanced sampling simulation at $\sim$ 10 GPa and 1400 K produced the same six-membered-ring molecule as in our unbiased simulations. Note that the free energy profile suggests that the cyclic form is in a metastable state, as shown in Fig. \ref{Fig:urail}(b).

Later we chose allophanic acid (C$_2$H$_4$N$_2$O$_3$, number 13 in Fig. \ref{Fig:chem_struc} ) and (Z)-ethene-1,2-diol (C$_2$H$_4$O$_2$, number 7 in Fig. S5) as reactants,
and built a new simulation box containing 5 C$_2$H$_4$N$_2$O$_3$, 5 C$_2$H$_4$O$_2$, and 10 H$_2$, as shown in Fig.\ref{Fig:urail} (c). At $\sim$ 10 GPa and 1400 K, a “C$_5$N$_2$” molecule, 3-(3-carboxyureido)-2,3-dihydroxypropanoic acid, 
was formed at around 28.8 ps.
This molecule has the same 
CN-backbone as the open-chain form of  thymine (T). As the simulation continued, a “C$_4$N$_2$” molecule, 2,3-dihydroxy-3-ureidopropanoic acid, labeled as “C$_4$N$_2$-2”, was formed, which has the same CN-backbone as the open-chain form of uracil. 
We further performed an enhanced sampling simulation to  analyze the ring formation process at $\sim$ 10 GPa and 1400 K. 
The simulation box has one “C$_4$N$_2$-2” molecule and 32 water molecules, as shown in Fig. \ref{Fig:urail} (c). Finally, a uracil-like molecule, called uracil glycol, was synthesized, which has a uracil ring.
The free energy curve in Fig. \ref{Fig:urail} (c) shows that the cyclic form is more stable than the open-chain form, indicating that the uracil-like molecule is possible to form in geological C-H-O-N fluids at extreme P-T conditions. 
The formation process and reaction mechanism of the uracil-like molecule are shown in Fig. S15. 
The uracil-like molecule can then serve as a precursor for the synthesis of other three types of nucleobases, providing the necessary components for forming RNA molecules.

Geochemical evidence of biotic activity suggests that the emergence of life can be traced back to as early as 3.8 billion years ago  \cite{mojzsis1996evidence}. 
The modern style of plate tectonics may have started about 3.2 billion years ago \cite{brenner2020paleomagnetic}, 
but some forms of plate tectonics may have already been active around 4.0 billion years ago,
and these tectonic activities might carry water, carbon, and other volatile substances down into the mantle \cite{korenaga2013initiation,HOLDER2024,blichert2008hafnium,hopkins2008low,huang2022barium}.
Because of reducing environments in early Earth's mantle \cite{aulbach2016evidence}, oxidized carbon and nitrogen can be reduced to CO and NH$_3$/NH$_4^+$, respectively (see Fig. S16)\cite{holm2009reduction}, 
and they may further synthesize to become key building blocks of life, which may reach the surface of early Earth.
Our findings also have important implications for the synthesis of organic molecules in interstellar space. 
The reactant molecules H$_2$O, H$_2$, CO and NH$_3$ largely exist in interstellar clouds and circumstellar shells, together with more than 200 complex organic species, such as aldehydes, alcohols, acids, amines and carboxamides \cite{guelin2022organic}.
The formation mechanism of those complex organic species may be intricate. 
Our  study suggests that the extreme P and T inside giant planets and their moons \cite{neri2020carbonaceous,nettelmann2016uranus}, like Jupiter and Saturn, could be considered as one important condition to help to synthesize some of those complex organic molecules \cite{postberg2018macromolecular}.
Earth's or other giant planets' interiors also provide a protective environment that
filters out harmful radiation and buffers against drastic physico-chemical variations.

Although our simulations were conducted under extreme P-T conditions, our study may also have implications for experiments at lower pressures and temperatures. We found that high pressures can facilitate the synthesis and stabilization of large biomolecules, aligning with some experimental studies conducted at lower pressures \cite{otake2011stability, ohara2007pressure, furukawa2012abiotic,  pedreira2019spontaneous}. It's important to note that some experiments were performed under water-poor conditions \cite{otake2011stability, ohara2007pressure, furukawa2012abiotic}, whereas we discovered that supercritical water plays a crucial role in the formation of biomolecules.

\section{Conclusions}
Extensive AIMD simulations have been performed to investigate the chemical reactions of small proto-organic molecules under the extreme P-T conditions. Three corresponding P-T systems were considered, \emph{i.e.}, 10 GPa - 1000 K, 10 GPa - 1400 K, and 13 GPa - 1400 K. The simulation results revealed the potential formation of substantial quantities of organic matter without any catalysts, with the composition and abundance of CN-containing species being profoundly influenced by the specific P-T conditions. Notably, formamide, recognized as a crucial precursor in the production of biomolecules, is observed to readily form under extreme P-T conditions, and its formation pathways are compared across different P-T conditions. Furthermore, free energy calculations confirmed that the 10 GPa - 1400 K condition is particularly favorable for the generation and preservation of CN-containing compounds. 

Using the molecules formed, we have further explored the formation pathways of large biomolecules that are directly related to the building blocks of life. Remarkably, glycine,  ribose, urea, and uracil-like molecules are successfully synthesized under extreme conditions in our AIMD simulations, thereby providing initial insights into the formation of organic compounds critical to the emergence of life. Particularly, different from that the six-membered ring form of ribose is the majority in solutions at ambient conditions, it was found that the formation of the five-membered ring form of ribose is more favored at extreme P-T conditions, which provides the necessary component for the generation of RNA molecules. 
Moreover, these compounds also have the potential to be transported to Earth's surface through geological activities, such as mantle convection, consequently contributing to the origin of life.
The complex organic species found in interstellar space may be also related to the chemical processes  inside giant planets and their moons at extreme P-T conditions.
Our comprehensive studies underscore the profound impact of extreme P-T conditions on the abiotic synthesis of organic matter and provide valuable insights into the mechanisms underlying the formation of crucial biomolecules for building blocks of life, which presents a novel perspective extending the cradle of life in planetary interiors. 

\section*{Computational Methodology}
\subsection{Ab initio molecular dynamics simulations (AIMD)}
We carried out AIMD simulations with the Born-Oppenheimer approximation using the Qbox code \cite{gygi2008architecture}.  
We applied the PBE exchange-correlation functional in density functional theory
\cite{perdew1996generalized}.
It is known that the PBE functional may be insufficient to describe aqueous solutions at ambient conditions, but our previous studies found that it works better at extreme P-T conditions than at ambient conditions \cite{pan2013dielectric, pan2016fate, stolte2019large, stolte2021water}, and many other studies also showed that this semi-local functional is reliable to study matter under extreme conditions \cite{spanu2011stability,cheng2023thermodynamics,zhang2023pbe}.
We used the ONCV pseudopotentials \cite{hamann2013optimized}, and the Bussi-Donadio-Parrinello (BDP) thermostat, with a relaxation time of 24.2 fs, to control the temperature \cite{bussi2007canonical}. To facilitate a larger time step of 0.24 fs, deuterium atoms were used in place of hydrogen atoms in our simulations. The simulation boxes have periodic boundary conditions. 
We first performed NPT simulations to achieve the desired pressures, configuring the plane-wave kinetic energy cutoff to 85 Ry and the cutoff for stress confinement potential to 70 Ry \cite{focher1994structural}.
We applied the simulation box sizes (Table S1) obtained from NPT simulations in the following NVT simulations for production runs, where the plane-wave kinetic energy cutoff was 65 Ry. 
During NVT simulations, we selected random time points to carry out independent simulations with the cutoff of 85 Ry to monitor pressure (Fig. S2).
For each P-T point, the NVT simulation time exceeds 400 ps (Fig. S3).
Our total AIMD simulation time exceeds 2.5 ns.

In our AIMD simulations, we calculated the forces on the nuclei using quantum mechanics, while treating nuclei including protons as classical particles. If we consider quantum nuclear effects, we anticipate an increase in the proton transfer rate, resulting in more active reactions involving proton transfer. However, at high temperatures, the significance of quantum effects may diminish compared to ambient conditions \cite{ceriotti2016nuclear}.

\subsection*{Free-energy calculations by coupling AIMD simulations with enhanced sampling}
To compute the free-energy landscape, we employed the adaptive biasing force method (ABF) \cite{darve2008adaptive} as implemented in the software package SSAGES \cite{sidky2018ssages} coupled with Qbox code. Unlike other advanced sampling methods that introduce bias potentials, this method aims to flatten the generalized force. 
The running average of the force in the $k$th bin along the reaction coordinate $\xi $ is calculated as:
\begin{align}
F_\xi(N_\text{step}, k) &= \frac{1}{\nu(N_\text{step}, k)}\sum_{i=1}^{\nu(N_\text{step}, k)} F_i(t_i^k) \\
F_i(t_i^k) &= \frac{d}{dt}\left(M_\xi\frac{d\xi}{dt}\right)\bigg|_{t_i^k}
\end{align}
where \(\nu \left( {{N}_{step}},k \right)\) is the number of samples collected in the $k$th bin after 
$N_{step}$ 
steps,  \({{F}_{i}}\left( t_{i}^{k} \right)\) is the $i$th force sample at the time \(t_{i}^{k}\), and M is the mass matrix. The bias force \(\text{-}\left( \nabla \xi  \right){{F}_{\xi }}\left( {{N}_{step,}}k \right)\) is added to the calculation until the free energy landscape is flat. 
We finally integrated the total biased force to get 
the free energy. For example, the free energy $\Delta {{A}_{a\to b}}$ between two states $a$ and $b$ is calculated as:
\begin{align}
\Delta {{A}_{a\to b}}=-\int_{{{\xi }_{a}}}^{{{\xi }_{b}}}{{{F}_{\xi }}d\xi \approx -\frac{{{\xi }_{b}}-{{\xi }_{a}}}{{{k}_{\max }}}}\sum\limits_{k=1}^{{{k}_{\max }}}{{{F}_{\xi }}}\left( {{N}_{step,}}k \right).
\end{align}
In our simulations, the calculated free energies are converged within 0.2 kcal/mol in the last 10 ps. The test of convergence of free energy calculations is shown in Fig. S17.

\section*{Supporting information}
Details of the simulations, Mole percents of C-N containing species over time, hydrocarbon products, free energy calculations, possible synthesis of ribose and uracil and their ring structures, compositions of fluids as a function of oxygen fugacity, test of convergence of free energy calculations.

\section*{Acknowledgements}
This work was supported by the Croucher Foundation through the Croucher Innovation Award, Hong Kong Research Grants Council (Projects GRF-16301723, GRF-16306621, and C6021-19EF), National Natural Science Foundation of China through the Excellent Young Scientists Fund (22022310).
R.T. would like to acknowledge support from the Original Exploration Plan-Key project (42150104).
Contributions by D.A.S. constitute material based upon work supported by the U.S. Department of Energy, Office of Science, Basic Energy Sciences, Geosciences program under Award Number DE-SC0019830 as well as NSF Petrology and Geochemistry Grant Number 2032039.
Part of this work was carried out using computational resources from the National Supercomputer Center in Guangzhou, China.

\bibliography{ref}

\clearpage

\begin{table}
\caption{Classification of the CN-containing species and the selected molecules.}
\label{Table 1}
\renewcommand*{\arraystretch}{1.0}
\centerline{
\begin{tabular}{c p{12cm}}
\hline 
\hline
     CN-containing   & Selected molecules \; \\
     \hline
“CN”   & CH$_2$NO, CH$_2$NO$_2$, CH$_3$NO, CH$_3$NO2, CH$_4$NO…               \\
“C$_2$N”  & C$_2$H$_2$NO$_2$, C$_2$H$_3$NO, C$_2$H$_3$NO$_2$, C$_2$H$_4$NO, C$_2$H$_4$NO$_3$, C$_2$H$_5$NO$_2$, C$_2$H$_5$NO$_3$…     \\
“CN$_2$”  &   CH$_4$N$_2$, CH$_6$N$_2$, CH$_7$N$_2$, CH$_4$N$_2$O…        \\
“C$_2$N$_2$”  &  C$_2$H$_2$N$_2$O$_3$, C$_2$H$_3$N$_2$O$_2$, C$_2$H$_3$N$_2$O$_3$, C$_2$H$_4$N$_2$O$_2$, C$_2$H$_4$N$_2$O$_3$, C$_2$H$_5$N$_2$, C$_2$H$_5$N$_2$O, C$_2$H$_5$N$_2$O$_2$, C$_2$H$_6$N$_2$O$_2$, C$_2$H$_7$N$_2$O$_2$…                \\
“C$_3$N$_x$”(“x = 1, 2, 3…”) & C$_3$H$_2$NO$_5$, C$_3$H$_3$NO$_3$, C$_3$H$_4$NO$_3$, C$_3$H$_4$NO$_4$, C$_3$H$_4$NO$_5$,
C$_3$H$_6$N$_2$O$_2$, C$_3$H$_7$N$_2$O$_2$, C$_3$H$_8$N$_2$O$_2$…  \\
“C$_4$N$_x$” &  C$_4$H$_3$NO$_5$, C$_4$H$_4$NO$_4$, C$_4$H$_4$NO$_5$, C$_4$H$_4$N$_2$O$_4$,
C$_4$H$_6$N$_2$O$_3$, C$_4$H$_7$N$_2$O$_4$…               \\
    \hline
    \hline
    \end{tabular}
    }
    \label{Tab_CN}
\end{table}

\end{document}